# Evidence of epigenetic oncogenesis: a turning point in cancer research

Jean-Pascal Capp [1], Benoît Aliaga [2] and Vera Pancaldi [2]

[1] Toulouse Biotechnology Institute, INSA / University of Toulouse, CNRS, INRAE, Toulouse, France

[2] CRCT, Université de Toulouse, Inserm, CNRS, Université Toulouse III-Paul Sabatier, Centre de Recherches en Cancérologie de Toulouse, Toulouse, France

Correspondence should be addressed to capp@insa-toulouse.fr and vera.pancaldi@inserm.fr

**Abstract**

In cancer research, the term epigenetics was used in the 1970s in its modern sense encompassing non-genetic events modifying the chromatin state, mainly to oppose the emerging oncogene paradigm. However, starting from the establishment of this prominent concept, the importance of these epigenetic phenomena in cancer rarely led to questioning the causal role of genetic alterations. Only in the last 10 years, the accumulation of problematic data, better experimental technologies, and some ambitious models pushed the idea that epigenetics could be at least as important as genetics in early oncogenesis. Until this year, a direct demonstration of epigenetic oncogenesis was still lacking. Now Parreno, Cavalli and colleagues, using a refined experimental model in the fruit fly *Drosophila melanogaster*, enforced the initiation of tumours solely by imposing a transient loss of Polycomb repression, leading to a purely epigenetic oncogenesis phenomenon. Despite a few caveats that we discuss, this pioneering work represents a major breakpoint in cancer research that leads us to consider the theoretical and conceptual implications on oncogenesis and to search for links between this artificial experimental model and naturally occurring processes, while revisiting cancer theories that were previously proposed as alternatives to the oncogene-centered paradigm.

**Introduction**

In 1935, Conrad Waddington suggested that cancers happen when cells have escaped from the action of normal growth-controlling agents, leading to a change in histological type and loss of their cellular identity [1]. According to his views, the developmental phenotype is the result of a complex, dynamic interaction between the developmental environment and the genome. To illustrate this complex phenomenon, he suggested a famous metaphor, the epigenetic landscape. It features a mountain leading to different valleys, in which differentiation of cells is represented by their rolling down into a specific valley, corresponding to a lineage. It is often used to pictorially describe the processes that we are discovering to be at the basis of cell differentiation and oncogenesis [2,3]. Actually, differentiation issues and changes in cellular identity have been noted for a very long time in cancer (the parallel between neoplasia and embryology is as old as the cell theory [4]). However, the molecular bases underlying these changes started to be deciphered only when the "modern" era of epigenetics began in 1975 [5].

**A short history of cancer epigenetics**

In 1975, Robin Holliday [6] and Arthur Riggs [7] independently proposed that DNA modification by methylation could play a major role in gene expression during cell differentiation and embryonic development. Simultaneously, chromatin studies proliferated, describing the chromatin state at the level of single genes. What we now call 'epigenetic inheritance' is a state of chromatin, distinct from the DNA sequence, which is transmitted during cell division (Box 1). Two main classes of epigenetic events are transmitted: methylation of DNA and several modifications of histones, collectively known as the epigenetic landscape. Over the past four decades, alterations of these phenomena have been implicated in carcinogenesis [8]. Apart from these heritable changes, chromatin remodeling, which is transient, is also important in oncogenesis [9,10]. Of note, both classes of events (heritable and transient) have been linked, particularly in pancreatic ductal adenocarcinoma (PDAC), where distinct epigenetic landscapes and nuclear architectures drive differentiation heterogeneity and allow PDAC classification into subtypes [11]. This potentially reveals an important structure-function principle that directly or indirectly relates the epigenome to changes in gene expression levels and phenotypes [12] associated to malignancy. Finally, other non-genetic events such as the transcription of coding and non-coding RNAs have been involved in cancer progression (Box 2) but not described as triggering factors.

Over 4 decades ago, it was already noted that loss of methylation plays an important role in activating gene expression. In 1983, this 'hypomethylation' was the first epigenetic defect to be described in cancer cells by Feinberg and Vogelstein [13]. In 2004, Feinberg proposed a chronology in which he began the history of cancer epigenetics with this study [14]. However, as early as 1971, Kauffman expressed doubts about the genetic dimension that was beginning to gain ground, especially through the theory of oncogenic viruses. He focused on tumor reversion experiments that enabled him to speak of 'epigenetic cancer' [15]. In 1979, Holliday proposed a new theory of cancer that drew on his work on epigenetics in earlier years [16]. In particular, he mentioned the fact that certain substances well identified as carcinogenic were not necessarily mutagenic. Despite these results, even if epigenetic changes in somatic cells were highlighted, they were thought to be mainly secondary effects of the repair of DNA lesions caused by carcinogenic substances, fostering his primarily genetic vision.

Similarly, the epigenetic research carried out by Feinberg and Vogelstein in the early 1980s accompanied the development of the oncogene paradigm [17], without distancing itself from it [13,18,19]. Prehn suggested in 1994 that cancer phenotypes would rather result from aberrant or abnormal expression of genes driven by epigenetic mechanisms (in the broad sense) than from mutations [20]. Thus, it might be more accurate to say that 'cancers beget mutations' than to say that 'mutations beget cancers' [15]. This viewpoint was contrary to the prevailing theory of somatic mutations and other genetic theories, which postulate that a 'renegade' cell possessing specific genetic alterations such as mutations in certain oncogenes or tumor suppressor genes is sufficient for cancerous development. It was not until 12 years later that Feinberg finally championed the idea of an epigenetic model competing with the genetic initiation model [21], as epigenetic changes in progenitor cells began to be seen as alternatives to mutations and chromosomal alterations in the disruption of gene function. . However, the question of genes remained central, as these precursor modifications were thought to be due to a new category of genes, the tumor progenitor genes, which were thought to be malfunctioning in early steps.

The link between genetic and epigenetic modifications was reinforced by the seminal discovery in the 2010s of mutations in chromatin remodelers that can be inactivated in a number of cancers [22]. However, these studies did not conceive epigenetic factors as initial and causal elements, as these deregulations are mutational in origin. Hanahan and Weinberg portrayed multistep tumor progression as a succession of clonal expansions, each of which is triggered by the chance acquisition of mutant genotypes [23]. Epigenetics was suggested to be one enabling characteristic that facilitates the acquisition of hallmark capabilities, especially because some of these mutations can be triggered by epigenetic changes affecting the regulation of gene expression. More recently, Hanahan [24] mentioned epigenetic dysregulation as a mechanism responsible for tumor heterogeneity without invoking its role in oncogenesis.

It was not until the second half of the 2010s that epigenetic alterations began to be considered in several models as an actual causal factor that could be at least as important as genetics for oncogenesis. It was first with the concept of constitutional epimutation [25], and then with the proposal that chromatin and epigenetic aberrations can confer the oncogenic properties necessary to induce tumorigenesis [26]. Genetic, environmental, and metabolic factors were suggested to make chromatin aberrantly permissive or restrictive. It was explicitly stated that 'purely epigenetic mechanisms may explain tumors that arise with few or no recurrent mutations' [26].

**Recent contributions of single-cell studies**

With the increasing availability of multiomics assays describing expression and chromatin accessibility, even in single-cells, studying the factors driving oncogenesis has become easier. In 2020, Marjanovic et al. [27] and Le Fave et al. [28] showed the emergence of a high plasticity state during lung cancer evolution in a genetically engineered mouse model using single-cell transcriptional and chromatin accessibility approaches, revealing important epigenomic alterations in early steps. They showed the loss of alveolar and lung-lineage identity as well as activation of Epithelial to Mesenchymal Transition (EMT) and embryonic endodermal and progenitor transcriptional profiles, thus identifying what they defined as a High Plasticity Cell State (HPCS) as a major transition point in tumour evolution. The HPCS showed high heterogeneity as well as an increase in proliferation potential and chemoresistance. A corresponding population was identified in human lung tumours. This and other examples show that tissue regeneration processes, in combination with oncogene activation, can lead to high plasticity appearing quickly at early stages of carcinogenesis.

More recently, Alonso Curbelo et al. [29] have studied similar mice models of pancreatic cancer showing that, in a KRAS mutated context, the presence of inflammation could divert tissue regeneration into malignancy. They also mechanistically proved that chromatin alterations on cell identity genes were involved in acinar to ductal metaplasia and the following inflammation induced de-differentiation that is part of normal tissue regeneration. They revealed a direct transition from acinar to neoplastic phenotypes and implicated a specific cytokine recapitulating the effect of injury as a trigger for pancreatic cancer development, thus reinforcing the importance of gene-environment interactions and epigenetics in early oncogenesis.

Furthermore, a single-cell analysis of these processes, coupled with innovative computational methods, revealed details about how the chromatin accessibility alterations in early neoplasia can lead to tumour development involving altered inter-cellular communication in the tissues [30]. Specifically, they observed that opening of chromatin in highly plastic states involved genomic regions that regulate genes important for communication, explaining a potential coupling between plasticity, inflammation and tissue remodeling that can lead to malignancy instead of regeneration.

Importantly, both these studies consider mice models that are produced by KRAS mutation, a well-known oncogene. The authors identified that this mutation produces high heterogeneity already in pre-malignant states, in some cases involving up-regulation of ZEB1 (associated with EMT) or of de-differentiation and developmental expression profiles, as well as oncogenic ones. The chromatin state in these different states could essentially prime some of these distinct cells already at early stages for later transition to malignancy via environmental triggers (inflammation).

**The first direct evidence of epigenetic oncogenesis**

While these works investigated specific cancer models in mice, a single-cell atlas of chromatin accessibility alterations from normal to malignant was recently produced in patient samples across 11 cancer types [31]. The data show epigenetic changes modifying chromatin accessibility and genetic mutations within the same pathway across cancer types, suggesting that certain changes in chromatin accessibility might represent critical events of cancer initiation [31]. However, the ultimate demonstration that cancer can be induced solely by epigenetic phenomena in absence of oncogenic mutations was still lacking until the recent work from Parreno *et al*. [32].

Since observing the initial stages of oncogenesis in mice models without oncogenic mutations still represents an enormous challenge, Parreno *et al.* considered a common model organism, the fruit fly *Drosophila melanogaster*, to study a potentially purely epigenetic oncogenesis phenomenon. They enforced the initiation of tumours solely by imposing a transient loss of Polycomb repression by inducible silencing of PRC1, removing a fundamental complex that represses developmentally regulated genes in differentiated cells [33]. Whereas most changes induced by this perturbation are transient, with de-repression of a considerable number of genomic regions, a few specific non-reversible changes are observed. These included activation of the ortholog of ZEB1 involved in EMT and of several components of the JAK-STAT pathway, potentially related to the de-repression of targets of the drosophila homolog of AP1. They also observed activation of the JNK signalling pathway, regulating response to external stimuli and bearing orthology relations to centralized stress response in yeast. These perturbations lead to the development of stable tumours.

Based on these results, the authors suggest a model according to which initial transient depletion of Polycomb repression leads to activation of the JAK-STAT pathway. This activation would mediate cell proliferation and the activation of an EMT process via the fly homolog of ZEB1, which prevents re-differentiation of the formed tumours. Since ZEB1 is known to induce EMT while stimulating DNA repair [34], one possibility is that its ectopic expression upon transient PRC1 depletion could contribute to tumorigenesis while preventing the accumulation of a high mutational burden. This might explain why Parreno and coworkers failed to detect increased mutation rates in their epigenetically initiated cancers. Interestingly, these tumours were also found to metastasize to distant regions, recapitulating a characteristic ability of tumours encountered in human cancers.

There are a few points that need to be highlighted before considering what these results might imply for naturally occurring cancers. First, there are obviously some key differences between mammals including humans and flies, one important one being the absence of DNA methylation in the model organism. Despite the conservation of several epigenomic factors across species, it is possible that the loss of such a major mechanism for genome control affects drosophila's ability to develop these tumours differently from what the scenario would be in mice or humans. Therefore, designing experiments that could produce at least initial evidence for (or against) the hypothesis that epigenetics can initiate cancer in mammals is required. Engineered systems allowing tissue-specific transient and reversible gene activation or inactivation via epigenetic editing [35] or RNA interference [36] could provide useful insight. Second, but not least important, this transient repression of Polycomb was obtained in a fairly artificial way. For these findings to have relevance in cancer, it would be fundamental to identify possible causes of a naturally occurring transient loss of Polycomb PRC1 complex functions (see below).

Despite these few caveats, the work from Parreno, Cavalli and colleagues represents a major breakpoint in cancer research that encourages us to consider the theoretical and conceptual implications on the definition of oncogenesis. In the following, we will discuss relevant cancer theories that were previously proposed as alternatives to the oncogene-centered paradigm.

**Other examples of studies questioning the oncogene paradigm**

After decades of dominance of the oncogene paradigm, and thanks to adaptations of the paradigm with discoveries about the role of epigenetics and the microenvironment, cancer researchers still rarely acknowledge that genetic alterations may not be the initiating events in cancer. However, recent years have seen the accumulation of evidence deeply questioning this paradigm, especially the massive presence of oncogenic alterations in normal tissues challenges the causal importance of these mutations [37]. The increasing evidence suggests the importance of 'additional' factors in oncogenesis [38], including environmental factors that would modulate clonal fate [39]. Thus, exposure to carcinogens such as cigarette smoke and pollutants, for instance, might favor the emergence of cancer as 'promoting' factors (see the well-known example on lung cancers by Swanton and colleagues [19]).

Recently, analyzing colorectal cancer samples at several levels at single-cell resolution showed a fairly weak correlation between phenotypic and genetic heterogeneity [40,41]. This suggests that the epigenome could indeed be at least as important as genetic factors in determining the strong phenotypic heterogeneity and plasticity associated with cancer. These analyses on independent colorectal cancer clones revealed clear patterns of mutations in chromatin modifier genes as well as chromatin accessibility alterations in regulatory regions of cancer driver genes. Interestingly, some of the epigenomic alterations involved potential activation of developmental genes via opening of regions bound by developmental regulators such as SOX and HOX family transcription factors (TFs). This is reminiscent of the alterations in Polycomb imposed in the Parreno *et al*. study. Furthermore, clonal alterations showed increased accessibility for JAK3 and in some cases regulation of SNAI TFs involved in EMT that were also identified in the fly epigenetic cancers described by Parreno *et al.* Finally, by comparing samples from the same or different tumours from the same patient, that include cells that are several cell divisions apart, they were also able to confirm that these epigenomic alterations are stable and heritable.

Several hypotheses have been proposed to explain the emergence of cancer through the modulation of the competitive fitness of mutant clones by environmental factors such as the metabolic context (closely linked to diet) or inflammation [42]. However, despite the 'missing links between clonal expansion and cancer transformation' [37], the necessary pre-existence of mutations has very rarely been called into question. Thus, it appeared for 10 years that somatic mutations are not sufficient to trigger oncogenesis without other alterations of non-genetic origin. Cavalli and colleagues' work now demonstrates that they are neither sufficient, nor necessary. Epigenetic events alone can be the triggering events.

**What can cause epigenetic alterations able to induce cancers?**

Considering the works cited above, we can imagine several scenarios that might lead to such a transient alteration of Polycomb repression. For instance, epigenetics-centered toxicological studies will be needed to provide new insights over the oncogenic role of non-mutagenic chemical substances in altering Polycomb regulation. Knowing the intimate relationships between metabolism and epigenetics [43], the role of dietary and metabolic changes will also of prime importance. More generally, an activation of stress responses in healthy tissues in the presence of strong inflammation leading to the activation of similar phenotypes as those described by Parreno *et al.* might be relevant. But more radically, the turning point that is the work of Parreno *et al.* should encourage us to consider the relationships between chromatin dynamics and tissue architecture and homeostasis.

Previous models and theories already suggested that disruption of tissue homeostasis alone can induce cancer. They are based for instance on works showing that genetic alterations in the hematopoietic stem cell niche can induce leukemia from healthy transplanted cells [44-46]. But once the tissue is disrupted, there is a disagreement over the underlying mechanisms.

A very radical position, known as the Tissue Organization Field Theory (TOFT) consists in proposing that the cell state by default is proliferation and that tissue organization allows control of this proliferation. It considers that the right level to explain the cancerous phenomenon is higher than the cellular one, namely that of biological organization and cellular communication in tissues. It posits that only the release of these tissue constraints can explain cancer [47]. In the TOFT, cell-centered processes are negligible and cannot explain cancer development. In this perspective, Parreno's works are hardly understandable, except if the epigenetic disorders caused by Polycomb inactivation create a massive disruption of the expression of proteins involved in cell interactions and communications. Of note, the results of Burdziak *et al.* [30] also pointed to the central role of ligands and receptor module expression in early carcinogenesis. It supported the importance of tissue organization and inter-cellular communication alterations, albeit in the context of an oncogenic mutation.

The pioneering studies of Mina Bissell on breast cancer highlighted the role of the extracellular matrix (ECM) and its interrelationships with the cells inside the tissue, as well as the carcinogenic importance of the matrix metalloproteinases (MMPs) that degrade the ECM. However, loss of adhesion caused by MMPs were proposed to attenuate DNA damage response and thus cause loss of genomic surveillance mechanisms and DNA damage-induced genomic instability [48]. Regarding the molecular pathways and mutations likely to cause disruption of breast tissue architecture, the role of mutated P53 was still emphasized [49]. Thus, a purely non-genetic view of cancer initiation was hardly privileged. Several experimental results pointed to an initiator/promoter model first proposed by Berenblum [50] and recently re-discovered [51] according to which it can be argued that healthy mutant cells are pre-existing and able to start malignant transformation in case of microenvironmental perturbation. These concepts have gained ground to explain how common carcinogens did not produce expected mutational signatures [52]. Nevertheless, it must be noted that this paradigm still allows us to consider genetic alterations as the ultimate origin of cancer [51].

Physical and biomechanical factors [53], which can play a role in changes to the extra-cellular matrix, and the disruption of cell-cell interactions [54], were also pointed out as causal events by proponents of the theory of attractors. Here, following such disruption, cells corresponding to healthy cell types and "normal" attractor states in the global gene regulatory network would be destabilized enough to reach pathological cancer attractors. These attractor states are not supposed to be reached during tissue development or renewal [55]. However, this network-centered view remains relatively elusive when searching for specific molecular events leading to such destabilization, warranting further studies.

Initial disruption of cell interactions and tissue homeostasis were also proposed to cause a global increase (from differentiated cells) or an uncontrolled maintenance (from adult stem cells) of cellular stochasticity [56-59]. The subsequent phenotypic plasticity would be due to the destabilized chromatin states and increased gene expression noise caused by dysregulated signaling pathways. This hypothesis is entirely compatible with the causative chain of events described above, linking mechanical or biochemical disruption in the microenvironment and the release of Polycomb from the chromatin. These events could potentially lead to a more widespread and stochastic gene expression.

Recently the main Polycomb component EZH2 was implicated in controlling lung fibrosis and epithelial remodeling after injury via a newly discovered transcriptional complex, which might constitute an alteration of Polycomb binding similar to that introduced by Parreno et al. [60]. An older study had identified EZH2 as key to preserving memory of mechanical perturbations in mesenchymal stem cells [61]. Mechanical impacts on transcriptional regulation were shown to impact Polycomb-mediated gene silencing. In melanoma, mechanical constraints during cell migration were shown to affect cellular plasticity and phenotypic switches via regulation of chromatin factors [62], with mechanisms which are reminiscent of neuronal development. Generally, there are strong links between cellular mechanical stimulation or environment and chromatin regulation [61]. This suggests that disruption of tissue homeostasis or mechanics could be sufficient to trigger Polycomb alterations driving epigenetic cancers.

**Conclusion**

In summary, the globally accepted viewpoint on oncogenesis is currently slowly shifting [63] and Parreno's study should accelerate this trend. It constitutes a turning point, even if we discussed that previous contributions in the field already suggested such a causal role of epigenetics. Future studies will be required to clarify the many ways by which initial epigenetic alterations can be acquired at a sufficient level to start transformation and the precise threshold needed. Investigations over the relationships with microenvironmental factors and the molecular consequences at the chromatin level that may destabilize cells enough to produce cell transformation without gene mutation will also help consolidating this new perspective on oncogenesis. While the estimation of carcinogenic potential of several substances is currently based on whether they induce DNA mutations, it might be necessary to also consider the effect of compounds on chromatin to capture potentially non-genetic carcinogens to improve cancer prevention. Moreover, the well-known effects of aging on the epigenome and chromatin structure, potentially affecting cellular plasticity, could help identify new modes of age-dependent increase in cancer risk, beyond the accumulation of DNA mutations. Finally, other non-genetic events beyond those directly affecting chromatin structure (such as metabolism or non-coding RNA production) should be considered in future studies as potential triggering factors in oncogenesis.

## Acknowledgements

V.P. acknowledges funding from the ARC foundation via the TRANSCAN SCIE-PANC project.

## Author contributions

J.-P.C., B.A. and V.P. collectively wrote the manuscript.

**Competing interests**

None.

**Data availability**

Not applicable as this is a "Recently in Press" article.

**Box 1: The different carriers of epigenetic information**

Since Waddington's definition of epigenetics, many publications have focused on its role in molecular biology, genetics, and development. Nowadays, epigenetic refers to a heritable, self-perpetuating, and reversible system [64]. Epigenetic regulation of gene expression will facilitate organisms' adaptation to changing environments. Today, there are three mechanisms of epigenetic information carriers: DNA methylation, post-translational modifications of histones, and non-coding RNAs (ncRNAs). These mechanisms are mediated by enzymes, which are classified into three groups: writers, readers, and erasers. Writers put an epigenetic mark on the histone tails or the DNA (5-methyl-cytosine) or on RNA ($m^6A$). For example, DNA methyltransferases (DNMTs) are enzymes that are capable of transferring a methyl group from the substrate S-adenosyl-L-methionine (AdoMet) to a 5-methyl-cytosine (5mC). Readers detect this epigenetic mark and contribute to the transcriptional machinery. The class of enzymes involved in DNA methylation is represented by the MBD (Methyl-CpG Binding Domain) family. Erasers remove these marks. For example, 5mC can be modified to thymine by AID/APOBEC enzymes, which will, in turn, be further modified to cytosine by the Thymine DNA Glycosylase mechanism.

**Box 2: Role of coding and non-coding RNAs in cancer**

RNA biology is a fast-evolving field, and recent decades have seen numerous breakthroughs in oncology. Firstly, it has been known since the 1950s that RNA molecules are decorated with modified nucleotides or RNA modifications. Nevertheless, the absence of methods to detect and map RNA modifications in low-abundance RNAs, along with the unclear functional significance of these modifications, had left the field of epitranscriptomics inactive for several decades [65]. This situation changed with the development of high-throughput sequencing. One of the most well-studied RNA modifications is $N^6$-methyladenosine ($m^6A$), which is the most prevalent modification found in eukaryotic mRNA. $m^6A$ has been implicated in cancer development and progression, with dysregulated $m^6A$ modification patterns associated with poor prognosis in several types of cancer (reviewed in [66]). Secondly, biologists initially thought that 90% of our genome was made up of "junk DNA," as much of it was transcribed into non-protein-coding RNA. This vision has progressively changed, revealing that most of the human genome is transcribed into RNAs that do not encode proteins (ncRNAs). In cancer, ncRNAs work as oncogenes or tumor suppressors to regulate carcinogenesis and progression. They influence cancer development by modulating various signaling pathways. These pathways include mitogen-activated protein kinase (MAPK), Wnt, Notch, and Hedgehog signaling. Additionally, ncRNAs impact Hippo signaling and the PI3K/AKT/mTOR pathway [67]. More interestingly, coding and ncRNAs interact together. Indeed, $m^6A$ regulatory factors participate in the regulation of tumors through ncRNAs through microRNA (miRNA), circRNA and lncRNAs [68]. On the other hand, ncRNAs (lncRNAs, miRNAs, and circRNAs) participate in the regulation of tumors through m6A regulatory factors [69].